\documentclass[a4paper,fleqn]{cas-dc}
\usepackage{threeparttable}

\def\tsc#1{\csdef{#1}{\textsc{\lowercase{#1}}\xspace}}
\tsc{WGM}
\tsc{QE}
\begin{document}
\let\WriteBookmarks\relax
\def\floatpagepagefraction{1}
\def\textpagefraction{.001}

\title [mode = title]{SurfGNN: A robust surface-based prediction model with interpretability for coactivation maps of spatial and cortical features}  

\author[1]{Zhuoshuo Li}

\affiliation[1]{organization={Department of Biomedical Engineering, Sun Yat-sen University},
            city={Shenzhen},
            postcode={518107}, 
            country={China}}
\affiliation[2]{organization={Institute of Biomedical Engineering, Ningbo Institute of Materials Technology and Engineering, Chinese Academy of Sciences}, 
	city={Ningbo},
	postcode={315201}, 
	country={China}}    

\affiliation[3]{organization={School of Cyber Science and Engineering, Ningbo University of Technology},
	city={Ningbo},
	postcode={315211}, 
	country={China}} 
\affiliation[4]{organization={Department of Radiology, School of Medicine, University of California San Francisco}, 
	city={San Francisco},
	state={CA},
	country={USA}}
\affiliation[5]{organization={USC Mark and Mary Stevens Neuroimaging and Informatics Institute, Keck School of Medicine of USC, University of Southern California},
	city={Los Angeles}, 
	state={CA},
	country={USA}}
\affiliation[6]{organization={Department of Radiology, Shenzhen  Maternity and Child Healthcare Hospital},
	city={Shenzhen},
	postcode={518028}, 
	country={China}}

\author[2]{Jiong Zhang}
\cormark[1]
\ead{jiong.zhang@ieee.org}   

\author[1]{Youbing Zeng}

\author[1]{Jiaying Lin}

\author[3]{Dan Zhang}

\author[1]{Jianjia Zhang}

\author[4]{Duan Xu}

\author[5]{Hosung Kim}

\author[6]{Bingguang Liu}
\cormark[1]
\ead{lbg379@163.com}
\author[1]{Mengting Liu}
\cormark[1]
\ead{liumt55@mail.sysu.edu.cn}

% Email id of the second author


\cortext[1]{Corresponding author}


\begin{abstract}
Current brain surface-based prediction models often overlook the variability of regional attributes at the cortical feature level. While graph neural networks (GNNs) excel at capturing regional differences, they encounter challenges when dealing with complex, high-density graph structures. In this work, we consider the cortical surface mesh as a sparse graph and propose an interpretable prediction model—Surface Graph Neural Network (SurfGNN). SurfGNN employs topology-sampling learning (TSL) and region-specific learning (RSL) structures to manage individual cortical features at both lower and higher scales of the surface mesh, effectively tackling the challenges posed by the overly abundant mesh nodes and addressing the issue of heterogeneity in cortical regions. Building on this, a novel score-weighted fusion (SWF) method is implemented to merge nodal representations associated with each cortical feature for prediction. We apply our model to a neonatal brain age prediction task using a dataset of harmonized MR images from 481 subjects (503 scans). SurfGNN outperforms all existing state-of-the-art methods, demonstrating an improvement of at least 9.0\%  and achieving a mean absolute error (MAE) of 0.827±0.056 in postmenstrual weeks. Furthermore, it generates feature-level activation maps, indicating its capability to identify robust regional variations in different morphometric contributions for prediction. 
\end{abstract}



%\nocite{*}

% Keywords
\begin{keywords}
brain surface prediction \sep cortical feature \sep GNN \sep spatio-cortical interpretability
\end{keywords}

\maketitle

% Main text
\section{Introduction}

There is growing evidence that brain development/aging trajectories and developments of brain disorders both could be traced on the cerebral cortex \cite{b1}. A prevalent approach for characterizing cerebral cortex is to reconstruct the cortical surface and measure the morphological features, such as cortical thickness, surface area, sulcal depth, myelin content, etc. \cite{b2} Currently, an important application of cortical features is to predict phenotypes, such as age \cite{b3,b4}, sex\cite{b5}, and brain disease states\cite{b6} using machine learning methods, and it could help explore important biomarkers about the cortex evolutional process and diagnose brain disorders.

For surface-based analysis, early approaches solely focused on vertex features without considering the topological structure of the surface mesh\cite{b8}. More recent techniques utilized GNN-based networks to collectively examine the node features and topological architecture of the cortical surface, due to the graph-like characteristics of the surface mesh\cite{b9}. However, managing cortical surface meshes that contain a vast number of vertices presents substantial computational difficulties in graph analysis. A common approach to address this is down-sampling the surface mesh, significantly reducing the vertex count before model learning \cite{b3,b4,b5}. Nevertheless, this could either diminish the prediction accuracy of the model or reduce its capacity to yield meaningful and interpretable results. 

Another commonly employed approach for cortical surface manipulation works within a spherical framework \cite{b10,b11,b12,b13}. It collects node features from the brain's surface, sequentially down-samples them, adhering to the hierarchical spherical architecture of the cortical surface, and eventually combines them for prediction. These models are efficient in managing large scale graphs with a high density of nodes. However, they couldn’t flexibly identify the best sub-graph structures or important nodes that contribute best to the prediction task. This might be important because different regions exhibit diverse responses to various predictive demands\cite{b17}. 

In addition to exploring the spatial heterogeneity at the global level, researches to date have seldom focused specifically on the feature-level heterogeneity. That is, different features may exhibit different spatial patterns in prediction tasks. Every cortical feature corresponds to a distinct macro- or micro- structure of the cerebral cortex. Observations during rapid developmental stages\cite{b17}, the aging process\cite{b20}, or under pathological conditions\cite{b21} have demonstrated that the regional variation in different cortical features could serve as distinct biomarkers for varying brain conditions. Enabling the separate manipulation of each cortical feature within the model, before integrating them for prediction analysis, might significantly boost the model's performance. We hypothesize that by allowing the autonomous expression of each feature, the model could capture more nuanced and impactful information for prediction tasks. It could help disaggregate the contribution of different cortical features to the prediction, enabling spatial-feature-level interpretations of the model for each subject.

A further challenge faced by surface-based models is their interpretability. It involves exploring the diverse characteristics of the cortical surface and identifying biomarkers associated with specific phenotypes. Post-hoc saliency-based methods are widely used, which aim to pinpoint the most impactful input features that contribute to a prediction task by examining the gradients or activations within the network in relation to a specific input\cite{b22}. Nevertheless, these methods have limitations\cite{b23,b24} and may not always hold for neuroimaging and neuroscience research, where available data are typically small-sized and much more complex \cite{b22,b25}. Another strategy is to develop a deep learning model with inherent self-interpretability\cite{b26}. This entails creating an end-to-end framework that facilitates the identification of detailed explanatory factors, thereby improving the extraction of discriminative representations and leading to more accurate outcomes. Previous research in this area, such as SiT\cite{b18} and NeuroExplainer\cite{b25}, has shown promising results. Motivated by these findings, creating a surface-based prediction model with built-in interpretability emerges as a significant and promising area of research.

To fulfill the outlined requirements and tackle the previously mentioned challenges, inspired by the GNN-based networks and the spherical frameworks, we propose a Surface Graph Neural Network (SurfGNN) as a self-interpretable prediction model. The entire framework of SurfGNN consists of topology-sampling learning (TSL) and region-specific learning (RSL) structures for each cortical feature, and a score-weighted fusion (SWF) structure across all features for prediction. We assess our model within the context of a brain age prediction task. Predicting brain age from structural brain neuroimaging data poses challenges akin to those faced in various neuroimaging applications, serving as a foundation for developing and testing deep learning algorithms. Furthermore, this task has gained attention due to its potential clinical and biological significance \cite{b57}. 

To summarize our contributions as follows:
\begin{enumerate}
\itemsep=0pt
\item We formulate a graph analysis process comprising TSL and RSL structures, extending from low-level to high-level surface meshes characterized by higher and fewer numbers of vertices, respectively. The TSL efficiently performs sampling on sparse graphs, preserving the overall brain topological shape, while the RSL effectively conducts in-depth graph analysis, distinguishing the varied impacts of different brain regions on prediction.
	
\item We propose a novel score-weighted fusion mechanism to amalgamate node information derived from individual cortical features within the graph learning framework. This mechanism also facilitates the generation of node scores, providing interpretable results that are specific to each feature.

\item We apply the SurfGNN to a neonatal brain age prediction task on a dataset with morphological features including cortical thickness, sulcal depth, and gray matter/white matter (GM/WM) intensity ratio. Our model outperforms state-of-the-art approaches on it. For each cortical feature, we build the spatial maps based on the node score.
\end{enumerate}

\section{Related works}

\subsection{Deep Learning Models on Cortical Surfaces}

Several strategies exist for applying deep learning models to the cortical surface in non-Euclidean spaces for predictive tasks. The first strategy involves obtaining the spectral domain and applying conventional convolutional neural networks (CNNs) on it \cite{b47,b8,b48}. It often performs poorly because models based on spectral features capture only global information and miss local details\cite{b49}. The second strategy involves projecting the original surface onto an intrinsic space, such as the tangent space, and then using 2D CNN models\cite{b50,b7}. This projection strategy, inescapably, introduces feature distortion and necessitates re-interpolation, complicating the network, which in turn elevates the computational burden and diminishes overall accuracy\cite{b10}. 

Over recent years, the spherical space has usually been applied to surface-based models in several studies. Zhao et al.\cite{b10} proposed the SphericalUNet with the Residual Hexagonal Convolution, while Monti et al.\cite{b14} used Gaussian MM Conv in their model MoNet. Jiang et al.\cite{b15} directly applied the MeshConv on the sphere with their model UG-SCNN. These models well preserve the complete surface structure during learning, but couldn’t flexibly identify the best subgraph structures or important nodes that contribute best to the prediction task. Additionally, Dahan et al.\cite{b18,b19} developed the application of vision transformers to spherical grid surfaces, implementing the self-attention mechanism. Nonetheless, the prediction accuracy of these models significantly relies on the selection of certain hyperparameters, like the shape and size of patches, which could be challenging to determine in practice.

Accordingly, in our work SurfGNN, we implement a graph neural network to formulate the feature learning process, efficiently managing complex non-Euclidean surfaces with a high density of nodes and exploring the regional heterogeneity of cortical surface for prediction. 

\subsection{Interpretable Methods}

Post-hoc methods are widely used in deep learning over gridded data, with some extensions to cortical surface cases. Besson et al.\cite{b5} employed the CAM method to map discriminative brain regions for both sex and age prediction tasks.  Liu et al.\cite{b4} applied perturbation methods with the surface-based GCN model for brain age prediction. Another approach involves building a deep learning model with self-interpretability, enhancing discriminative representation extraction and ultimately yielding precise results. Xue et al.\cite{b25} proposed a NeuroExplainer model based on an attention mechanism for a classification task of identifying preterm infants. Dahan et al.\cite{b18} utilize the self-attention module of the vision transformer model to construct attention maps for brain age prediction. Additionally, Li et al.\cite{b37} utilized the score produced by the distinct node drop pooling structure in the GNN to elucidate the signification of nodes.

Inspired by that, we employ a pooling-like approach to generate node scores, as attention weights to facilitate weighted fusion among sparse graphs of each cortical feature. It is imperative to highlight that our model endows the generated node scores with the capability to not only assess the regions of a single cortical feature but also facilitate comparisons across different features.

\section{Method}

\begin{figure*}[!t]
	\centering
	\includegraphics[width=0.8\textwidth]{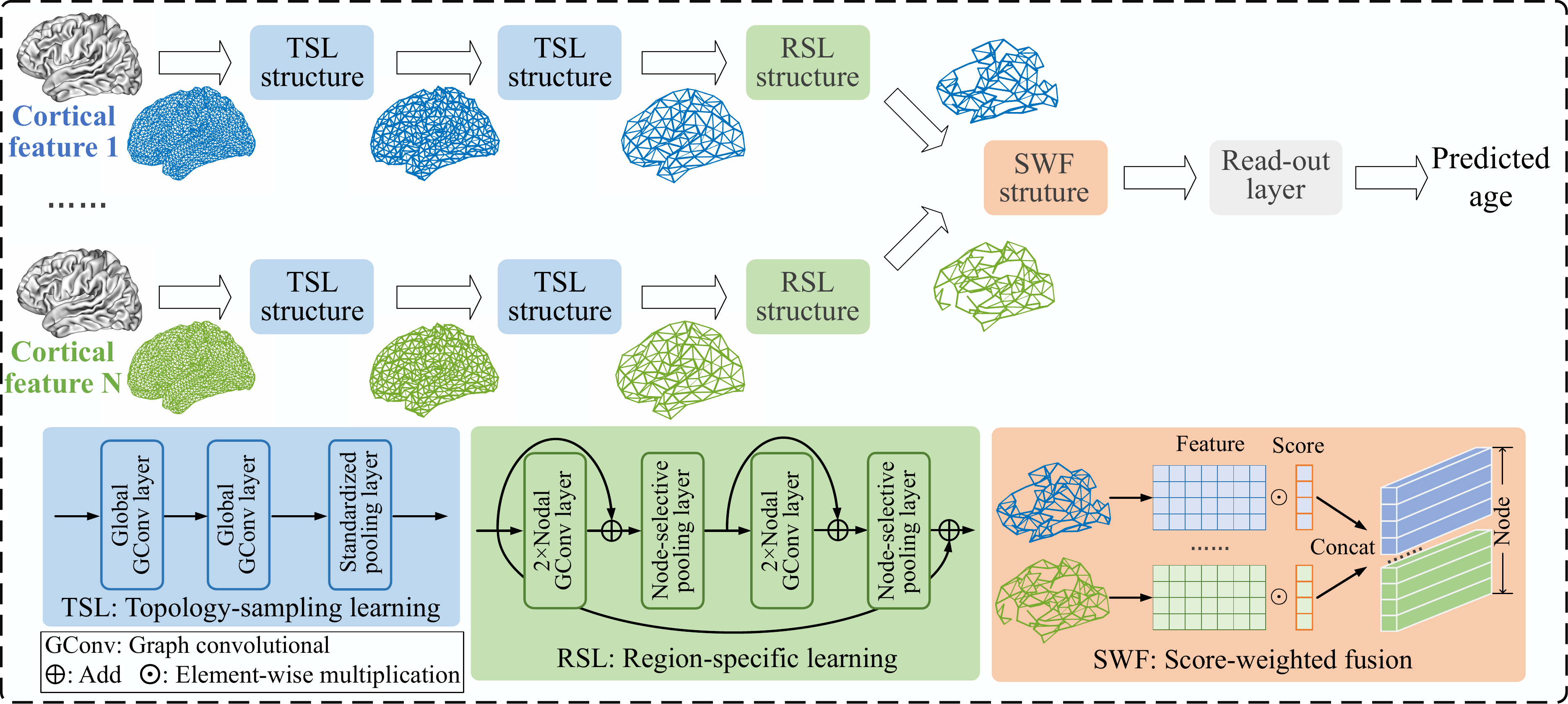}
	\caption{Overview of the proposed network architecture and its key modules. N: the number of cortical features for each subject, input into SurfGNN. The model showcased above operates on an input mesh resolution of 5,124 nodes, necessitating two TSL structures for each cortical feature. For node number of 81,924, 20,484, 1,284, the model requires four, three and one TSL structures, respectively.}
	\label{fig1}
\end{figure*}

Fig. \ref{fig1} illustrates the complete architecture of SurfGNN. We first outline the approach for extracting cortical surfaces and morphological features from brain MR images. Then we provide a comprehensive exposition of each module within SurfGNN. This includes specific graph convolutional layers and graph pooling layers, utilized in the topology-sampling learning and the region-specific learning structures, along with the score-weighted fusion structure incorporating a read-out layer to predict phenotypes. Additionally, we describe the employed loss functions during the network training process.

\begin{figure}[!t]
	\centering
	\includegraphics[width=\columnwidth]{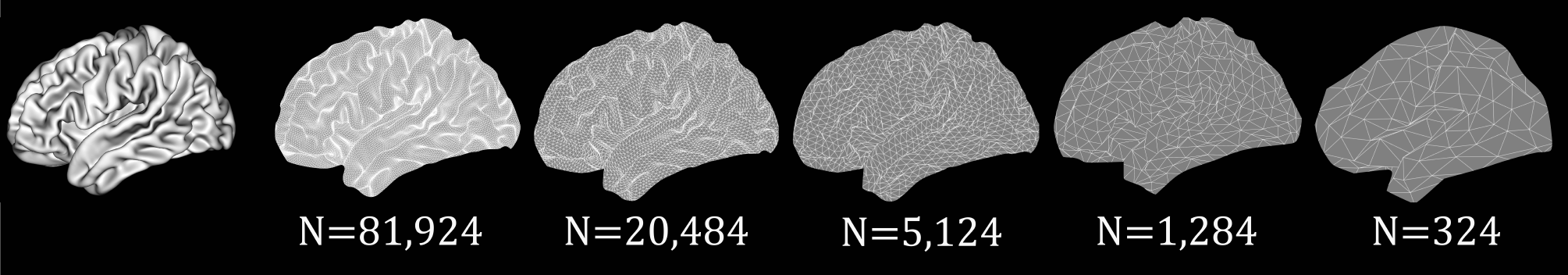}
	\caption{Multi-resolution mesh maps on cerebral cortical surface. N: the number of nodes on the mesh.}
	\label{fig2}
\end{figure}

\subsection{Surface Reconstruction and Morphological Features Extraction}

Brain cortical features input to the model are retrieved from T1-weighted MR images, through NEOCIVET\cite{b27,b28,b29}, a pipeline for neonatal brain MRI processing. Beginning with general MR image preprocessing—comprising denoising, intensity nonuniformity correction, and brain extraction—the process includes brain tissue segmentation to delineate distinct regions such as white matter (WM), graph matter (GM) and CSF. A marching-cube-based framework generates a triangulated mesh representing the WM surface along the GM-WM boundary. The surface mesh is resampled to a fixed number of 81,924 vertices using the icosahedron spherical fitting, then refined to obtain the CSF surface along the sharp edge of the GM-pial interface. Cortical features, including cortical thickness, sulcal depth, and GM/WM intensity ratio, are measured at each vertex on the surface. Cortical surfaces with these features are later transformed into a consistent template using the transformation obtained in the surface registration, enabling inter-subject comparisons. 

Following the pipeline, a surface mesh, covering both WM and CSF surfaces for each brain, is generated, containing 81,924 vertices and cortical feature measured at each vertex. Each surface mesh, with its 81,924 vertices, then undergoes down-sampling using the icosahedron down-sampling approach\cite{b30} to four additional levels of resolution, specifically 20,484, 5,124, 1,284, and 324 vertices, as Fig.\ref{fig2} shows. Notably, at lower resolutions, such as 84 vertices, the mesh struggles to delineate a distinct brain structure. Thus, a mesh with 324 vertices is typically considered the minimum resolution with the NEOCIVET pipeline.

Above all, the surface mesh of the brain is modeled as a sparse graph, where surface vertices serve as nodes, cortical features as node features, and the surface mesh as edges in the sparse graph. Edge features are represented as binary values (0 or 1), indicating whether the two vertices are connected or not in the structure mesh.

\subsection{Topology-sampling Learning Structure}

The model processes complex sparse graphs derived from high-node-count surface meshes. To manage this complexity while preserving the original brain shape, we introduce the topology-sampling learning (TSL) structure, which inspired by the previous spherical framework \cite{b7, b10}. It is utilized to integrate node context and to systematically sample sparse graphs according to the surface order illustrated in Fig. \ref{fig2}. 

\subsubsection{Global Graph Convolutional Layer}

The purpose of the graph convolutional layers here is to extract low-dimensional surface information from the graph. A common approach models is the 1-ring hexagonal convolution applied in spherical models \cite{b10}. However, unlike in spherical meshes, our irregular surface meshes face inconsistencies in adjacent node counts. To address this, our improvement strategy based on the hexagonal convolution, involves utilizing maximum number of adjacent nodes and representing the features of absent nodes using zero, giving rise to a novel graph convolutional layer—global graph convolutional layer.

For a sparse graph with $ N^{(l)} $ nodes in the $ l^{th} $ layer, $ v_i^{(l)} $ represents the $ i^{th} $ node and $ x_i^{(l)} \in \mathbb{R}^{d^{(l)}}$ as its feature. $ E_i^{(l)} $ is defined as the number of $ v_i^{(l)} $'s 1-hop neighbors plus one (the one represents the node $ v_i^{(l)} $ itself). This value represents the number of nodes within a one-edge distance from $ v_i^{(l)} $ and is used to determine the filter size for convolutional layers. Since different nodes may have varying numbers of neighbors, $ E_i^{(l)} $ is not consistent across all nodes. To maintain efficiency, we use the maximum value of $ E_i^{(l)} $, denoted as $ E^{(l)} $, as the size of the convolutional filter. Moreover, $ \widetilde{x}_i^{(l)}=[x_i^{(l)},\cdots,x_{E_i^{(l)}}^{(l)}] $ represents the feature of $ v_i^{(l)} $ to learn. In cases where $ E_i^{(l)}<E^{(l)} $, we employ zeros to complete the features of absent neighboring nodes and it must be $ \widetilde{x}_i^{(l)} \in \mathbb{R}^{1 \times E^{(l)}d^{(l)}} $. Notably, the order of nodes is based on the angle between the vector of center vertex to neighboring vertex and the x-axis in the tangent plane. If there are not enough neighboring nodes, the zero-padding is applied at the end. 

Ultimately, we obtain the feature matrix. $ \widetilde{X}^{(l)}=\\{{[\widetilde{x}_1^{(l)},\cdots,\widetilde{x}_{N^{(l)}}^{(l)}]}^\top} $ as the input for the convolutional layer across the entire graph. The filter $ W \in \mathbb{R}^{E^{(l)}d^{(l)} \times d^{(l+1)}} $ is established to learn it, and the output graphs with the same $ N^{(l)} $ nodes and $ d^{(l+1)} $ feature channels can be obtained.

The zero-padding strategy is employed to maintain consistency in global node information transmission, ensuring that each node fully and uniquely integrates the information from its adjacent nodes. Notably, since the number of nodes requiring zero-padding is minimal, its impact on the convolution direction is negligible.

\subsubsection{Topology-preserved Pooling Layer}

The pooling layer in this instance is designed to merge local features and transmit them to upper levers. Whereas it is similar with the node clustering pooling method\cite{b31} in typical GNN models, a key distinction lies in the uniform and consistent pattern of node clustering across all nodes from different subjects for topology preservation, which is based on the organization of vertices on each scale of meshes, as shown in Fig. \ref{fig2}. This pooling strategy is widely employed in the spherical framework \cite{b7, b10}.

Specifically, the pooling operation for node $ v $ involves employing the average pooling operation to combine its adjacent nodes feature, and then, nodes and features corresponding to a sparse graph at the next resolution are preserved. That is, assuming the graph before pooling with a count $ N^{(l)} $ of 81,924 nodes, after a single round of pooling, the node count $ N^{(l+1)} $ reduces to 20,484 of the output graph, following the formula $ N^{(l+1)}=(N^{(l)}+12)/4 $. This formula is derived from the process of triangularizing the surface mesh during the construction of brain surfaces.

\subsection{Region-specific Learning Structure}

Following the previously TSL structure, wherein graph information across the surface is compiled and transferred to the upper level of the surface mesh, we introduce the region-specific learning (RSL) structure for deep graph analysis and explore the regional contributions towards with the prediction. 

\subsubsection{Nodal Graph Convolutional Layer}

The graph convolutional layer is used to learn the graph information with the extraction of node features in high dimension. 

For the graph of each subject, we define $ v_i $ as the $ i^{th} $ node, $ x_i^{(l)} \in \mathbb{R}^{d^{(l)}} $ as features for node $ v_i $  in the $ l^{th} $ layer. The forward propagation of node feature can be calculated as:
\begin{equation}
	\widetilde{x}_i^{(l+1)}=\mathrm{relu}(w_i^{(l)}x_i^{(l)}+\sum\nolimits_{j \in \mathcal{N}^{(l)}(i)}w_j^{(l)}x_j^{(l)})
	\label{eq1}
\end{equation}
where $ w_i^{(l)} $ is the embedding kernel to be learned for the $ i^{th} $ node in the $ l^{th} $ layer, $ \mathcal{N}^{(l)}(i) $ denotes the set of indices of neighboring nodes of node $ v_i $. Eq. \ref{eq1} illustrates the message passing mechanism \cite{b56} in the graph convolution process, , inspired by Li et al. \cite{b37} With the embedding kernel $ w^{(l)} $ to update each node’s message, the message of all neighbors $ \mathcal{N}^{(l)}(i) $ to node $ v_i^{(l)} $are aggregated by summation, which facilitates the message passing to node $ v_i^{(l)} $. During this process, the message of node $ v_i^{(l)} $ is retained to prevent the loss of valuable information.
%Fig. 4 shows the calculation process of \eqref{eq1}. 

In contrast to typical graph convolution, the embedding kernel $ w_j^{(l)} $ belongs to the $ j^{th} $ node, rather than the mapping from the node $ v_i $ to node $ v_j $. Thus, what we need to learn is the weight matrix as the embedding kernel $ W^{(l)}=[w_1^{(l)},\cdots,w_{N ^{(l)}}^{(l)}]\top $ for $ N^{(l)} $ nodes on the entire graph indicating the mapping from the $ l^{th} $ layer to the $ (l+1)^{th} $ layer. A multi-layer perceptron ($ \mathrm{MLP} $) with two basic functions is employed to conduct it as follows:
\begin{equation}
	W^{(l)}=\mathrm{MLP}^{(l)}(\mathbf{r})=\mathbf{m}_2^{(l)}\mathrm{relu}(\mathbf{m}_1^{(l)}\mathbf{r})+\mathbf{b}^{(l)}
	\label{eq2}
\end{equation}
where the $ \mathbf{r} $ is a position matrix with one-hot encoder, and the parameters $ \left\{\mathbf{m}_1^{(l)},\mathbf{m}_2^{(l)},\mathbf{b}^{(l)}\right\} $ help to generates a $ d^{(l+1)} \times d^{(l)} $ vector for each node. We denote a middle dimension $ k^{(l)} $, and then the number of parameters for $ \mathbf{m}_1^{(l)} $ and $ \mathbf{m}_2^{(l)} $ are respectively  $ k^{(l)} \times N^{(l)} $ and $ d^{(l)}d^{(l+1)} \times k^{(l)} $. Compared with directly generating a size of $ d^{(l)}d^{(l+1)}N^{(l)} $ for $ W^{(l)} $, the $ \mathrm{MLP} $ operation significantly reduces the quantity of learnable parameters where $  k^{(l)}\ll N^{(l)} $.

\subsubsection{Node-selective Pooling Layer}

The node-selective pooling layer here is aimed to highlight the significant nodes by removing unnecessary ones. The selection criterion hinges on the principle that, if a node’s feature can be adequately reconstructed by its adjacent nodes, its removal would lead to minimal loss of pertinent information for the entire graph. According to this, we calculate the node score $ p^{(l)} $ representing the criterion by utilizing the Manhattan distance between each node and its neighbors\cite{b32}, as follows:
\begin{equation}
	p^{(l)}=\Vert (I^{(l)})-(D^{(l)})^{-1}A^{(l)}\widetilde{X}^{(l+1)} \Vert_1
	\label{eq3}
\end{equation}
where $ \widetilde{X}^{(l+1)}=[\widetilde{x}_1^{(l+1)},\cdots,\widetilde{x}_{N^{(l)}}^{(l+1)}]\top $ denotes the node feature matrix from the nodal graph convolutional layer, and the $ A^{(l)} \in \mathbb{R}^{N^{(l)} \times N^{(l)}} $ is the adjacent matrix. $ D^{(l)} $ represents the diagonal matrix of $ A^{(l)} $ and $ I^{(l)} $ is the identity matrix. The $ \Vert\cdot\Vert_1 $ performs $ \ell_1 $ norm row-wisely.

We thereby obtain the score $ p^{(l)} $ for each node on the graph. Notably, a node with a substantial score signifies a potentially significant information gap between it with its neighbors, therefore necessitating retention. Conversely, if the node score is low, it implies effective re-constructability and suitability for removal. $ N^{(l+1)} $ nodes for the output graph after once pooling are reserved and the sparse graph would be updated as follows:
\begin{equation}
	\begin{split}
		N^{(l+1)}&=N^{(l)} \times r, \ \ \ \ \ \ \  \mathrm{idx}=\mathrm{top}(p^{(l)}, N^{(l+1)}), \\
		X^{(l+1)}&=\widetilde{X}^{(l+1)}(\mathrm{idx},:),\  A^{(l+1)}=A^{(l)}(\mathrm{idx}, \mathrm{idx})
	\end{split}
	\label{eq4}
\end{equation}
where $ r $ is defined as the pooling ratio and $ \mathrm{top} $ operation returns the index of the $ N^{(l+1)} $ largest values in $ p^{(l)} $. Then these $ N^{(l+1)} $ nodes with the feature matrix $ X^{(l+1)} $ and adjacency matrix $ A^{(l+1)} $ could form a new sparse graph as the output of pooling layer. We set the parameter $ r $ as 0.75 for all node-selective pooling layers.

The RSL structure consists of four nodal graph convolutional layers and two node-selective pooling layers for each cortical feature in our SurfGNN. Considering that graph convolutional layers do not scale well to deep architectures, which may  introduce a high complexity of backpropagation and over-smoothing issues, we integrate a residual module into this structure, inspired by the ResGCN\cite{b33}, as Fig. \ref{fig1} shows. 

\subsection{Score-weighted Fusion Structure}

An independent feature extraction model comprising TSL and RSL structures, are utilized to handle the input of each sparse graph representing a cortical feature. Subsequently, a fusion strategy becomes imperative to combine these diverse graphs for predictive modeling.

We define $ C $ as the number of graphs as well as cortical features. $ N^j $ denotes the number of the remaining nodes on the $ j^{th} $ graph and $ x_i^j $ denotes the feature of node $ v_i^j $ . We concatenate the feature of all the nodes from each graph, denoted as $ X=[x_1^1,\cdots,x_{N^1}^1,\cdots,x_1^j,\cdots,x_{N^j}^j,\cdots,x_1^C,\cdots,x_{N^C}^C]\top $.

Inspired by the graph drop pooling\cite{b31}, we propose a novel score-weighted fusion (SWF) strategy. That is, to assign a score to each node before concatenation, indicating its significance for prediction. To accomplish this, all node features are projected onto a trainable vector $ \mathbf{w} $ to generate the respective scores $ \mathbf{s} $, and subsequently updated through multiplication with them. The calculation procedures are as follows:
\begin{equation}
	\mathbf{s}=\mathrm{sigmoid}(X\mathbf{w}/\Vert\mathbf{w}\Vert_2),\ \ \  \widetilde{X}=X \odot \mathbf{s}
	\label{eq6}
\end{equation}
where the $ \Vert\cdot\Vert_2 $ is the $ \ell_2 $ norm and $ \odot $ denotes the element-wise matrix multiplication.

Projecting all node features onto a common vector enables the direct comparisons of resulting node scores. Besides, weighting these scores into the features facilitates a feedback process that iteratively learns and updates scores, ensuring their reliability. Moreover, these derived scores encompass all remaining nodes from each graph, enabling the generation of activation maps for each cortical feature input and enhancing the model's interpretability. The structure of SWF can be seen in Fig. \ref{fig1}.

Lastly, a read-out layer is employed to predict the phenotype. We use the global average pooling (GAP) and global max pooling (GMP) to integrate all nodes, as follows:
\begin{equation}
	z=\mathrm{GAP}(\widetilde{X})\Vert\mathrm{GMP}(\widetilde{X})
	\label{eq7}
\end{equation}
where the $ \Vert $ denotes concatenation. The final vector $ z $ is then sent to the regressor composed of two fully connected layers to obtain the prediction results. Global pooling layers are utilized to obtain a comprehensive and abundant global graph representation for the subsequent regressor. Additionally, compared to using the regressor without global pooling layers, this approach effectively reduces the number pf parameters for fully connected layers, thereby alleviating overfitting.

\subsection{Loss Functions}

\subsubsection{Regression Loss}

During the network training procedure, we adopt the mean squared error (MSE). It is calculated as:
\begin{equation}
	L_\mathrm{MSE}=\frac{1}{M}\sum_{m=1}^{M}(y_m-\hat{y}_m)^2.
	\label{8}
\end{equation}
Here $ M $ is the number of instances, $ y_m $ is the known PMA of the $ m^{th} $ subject and $ \hat{y}_m $ is the predicted age from the networks.
\subsubsection{Feature Loss}

Since cortical features undergo individualized information extraction processes, a considerable number of learnable parameters are introduced. To enhance prediction accuracy, we propose the concept of feature loss, which imposes additional learning constraints. The feature loss avoids reliance on a single cortical feature and ensures that information from all features is efficiently learned. In detail, additional readout layers are employed after each RSL structure, to generate individual prediction outputs from each cortical feature. These outputs are used to calculate the MSE, forming the feature loss. It is worth noting that they are solely for loss calculation and do not represent the final prediction results. Consequently, they are not shown in Fig. 1 to avoid any misunderstanding. The feature loss is calculated as follows:
\begin{equation}
	L_\mathrm{feature}=\sum_{c=1}^{C}\frac{1}{M}\sum_{m=1}^{M}(y_m-\hat{y}_{m,c})^2.``
	\label{eq9}
\end{equation}
Here C denotes the number of cortical features in the model and $ \hat{y}_{m,c} $ denotes the output from $ c^{th} $ features.
\subsubsection{Unit Loss}

For the learnable vector $ \mathbf{w} $ in the SWF structure, it may cause an identifiability issue, i.e. $ \mathbf{w} $ could be arbitrarily scaled but it has no effect on the calculation of scores in \eqref{eq6} and generates the same scores. Thus, we adopt a unit loss to constraint $ \mathbf{w} $ to be a unit vector:
\begin{equation}
	L_\mathrm{unit}=(\Vert\mathbf{w}\Vert_2-1)^2.
	\label{eq10}
\end{equation}
\subsubsection{Score Loss}

To accentuate the differentiation among nodes in the SWF structure and promote increased sparsity within score maps, we adopt a score loss to expand the distribution of the scores. Sort the scores $ \mathbf{s} $ in descending score, and the score loss is defined using binary cross-entropy as:
\begin{equation}
	L_\mathrm{score}=-\frac{1}{M}\sum_{m=1}^{M}\frac{1}{N}(\sum_{i=1}^{0.5N}\log(\mathbf{s}_{m,i})+\sum_{i=1}^{0.5N}\log(1-\mathbf{s}_{m,i+0.5N}))
	\label{eq11}
\end{equation}
where N denotes the number of scores, e.g. the number of the remaining nodes for each subject, and $ s_{m,i} $ means the score of the $ i^{th} $ node for the $ m^{th} $ subject.

Finally, the total training loss is defined as:
\begin{equation}
L_\mathrm{total}=L_\mathrm{MSE}+\lambda_\mathrm{1}L_\mathrm{feature}+\lambda_\mathrm{2}L_\mathrm{unit}+\lambda_\mathrm{3} L_\mathrm{score}
\label{eq12}
\end{equation}
where $\lambda_\mathrm{1}$, $\lambda_\mathrm{2}$ and $\lambda_\mathrm{3}$ are the tunable hyperparameters to balance weights of losses. Based on hyperparameter selection experiments, we set $\lambda_\mathrm{1}$ = 1, $\lambda_\mathrm{2}$ = 1 and $\lambda_\mathrm{3}$ = 0.1 for the subsequent experiments to achieve optimal prediction performance.

\subsection{SurfGNN Architecture}
In summary, our SurfGNN model adopts a multi-graph input framework, where each graph corresponds to a specific cortical feature. It allows sparse graphs across various resolutions—81,924, 20,484, 5,124, and 1,284 nodes—as input. A sparse graph with a consistent structure across diverse cortical features is generated via TSL structures. It then advances through RSL structures. Finally, the proposed SWF mechanism combines the learning results from each cortical feature, followed by the utilization of a read-out layer for prediction. Fig. \ref{fig1} presents the process designed for the input surface with 5,124 nodes and the graph featuring 324 nodes between TSL and RSL structures, which is the main architecture through our experiments. Also, we will discuss the influence with different input or output resolutions with our model, which would be specifically noted.

\section{Experiments}
\subsection{Datasets}
Our dataset consist of T1w MR images from 92 preterm neonates admitted to UCSF Benioff Children’s Hospital San Francisco and 389 neonates from the developing Human Connectome Project (dHCP)\cite{b34,b35}. Since most subjects from UCSF were scanned twice, with some excluded due to poor quality of imaging, the dataset does contain a total of 503 MRI scans. Notably, the MRIs have been visually inspected and excluded those with major focal parenchymal lesions. Table \ref{table1} presents the extensive information about the demographic characteristics. MR images of both cohorts have undergone the preprocessing process outlined in Section 3.1 to extract the cortical surface meshes and morphological features (cortical thickness, sulcus depth and GM/WM intensity ratio), which are used to predict the postmenstrual age. Given that the potential deviation features brought on by different field strengths of MRI scanners and different sites, we adopt the Combat function\cite{b36} to harmonize all the cortical features.

\begin{table}[]
	\caption{The information of datasets used for pma prediction.}
	\renewcommand{\arraystretch}{1.1}
	\centering
	\begin{tabular}{cccc}
		\toprule[1pt]
		Dataset                                                                & UCSF      & The dHCP  & Total     \\ \hline
		Subjects (n)                                                           & 92        & 389       & 481       \\
		MRI scans (n)                                                          & 114       & 389       & 503       \\
		\begin{tabular}[c]{@{}c@{}}PMA at scan\\ (weeks, mean$ \pm $std)\end{tabular} & 33.83$ \pm $2.86 & 39.76$ \pm $2.98 & 38.42$ \pm $3.86   \\
		\bottomrule[1pt]
	\end{tabular}
	\label{table1}
\end{table}

\subsection{Training Details}
In our implementation of SurfGNN in Fig.\ref{fig2}, the feature representations of the four convolutional layers in the two TSLs and the four layers in the RSL have 16, 16, 32, 32, 128, 128, 256 and 256 channels, respectively. We train and test the SurfGNN model on Pytorch in the Python environment using an NVIDIA GeForce RTX 3090 with 24 GB GPU memory. The total training epochs is 300, with an initial learning rate of 0.001 and annealed to half every 50 epochs, and Adam is used as an optimizer.
The merged dataset is randomly split into 5 folds of approximately equal sample sizes without overlapping and we employ a 5-fold cross-validation strategy to evaluate the performance of SurfGNN. Importantly, to prevent data leakage, scans from the same subject are consistently placed within the same fold. The accuracy metrics provide an evaluation of a model’s prediction performance: mean absolute error (MAE) and Person correlation coefficient (PCC).

\section{Results}
\subsection{Comparison with the State-of-the-art Methods}
We compare the proposed SurfGNN model with the following four different state-of-the-art methods. (1) SphericalUNet. A variation of its original architecture\cite{b10} for regression tasks by removing the up-sampling decoder. (2) MoNet\cite{b14}. The mixture model network utilizes the Gaussian mixture model convolutional operators to replace filters in SphericalUNet. (3) SiT\cite{b18}. The surface vision transformer treats the surface as a sequence of triangular patches and encodes them with the transformer model. (4) BrainGNN\cite{b37}. A GNN model on brain data with the roi-aware graph convolutional layers and the roi-topk pooling layers. 

In our experiments, we utilized sparse surfaces at various resolutions (as illustrated in Fig. \ref{fig2}) as inputs for the comparison models.  It's important to note that sparse graphs featuring 81,924 nodes could not be utilized with BrainGNN due to GPU memory limitations. The above methods are implemented based on the released codes with our best efforts.

\begin{figure}[!t]
	\centering
	\includegraphics[width=0.7\columnwidth]{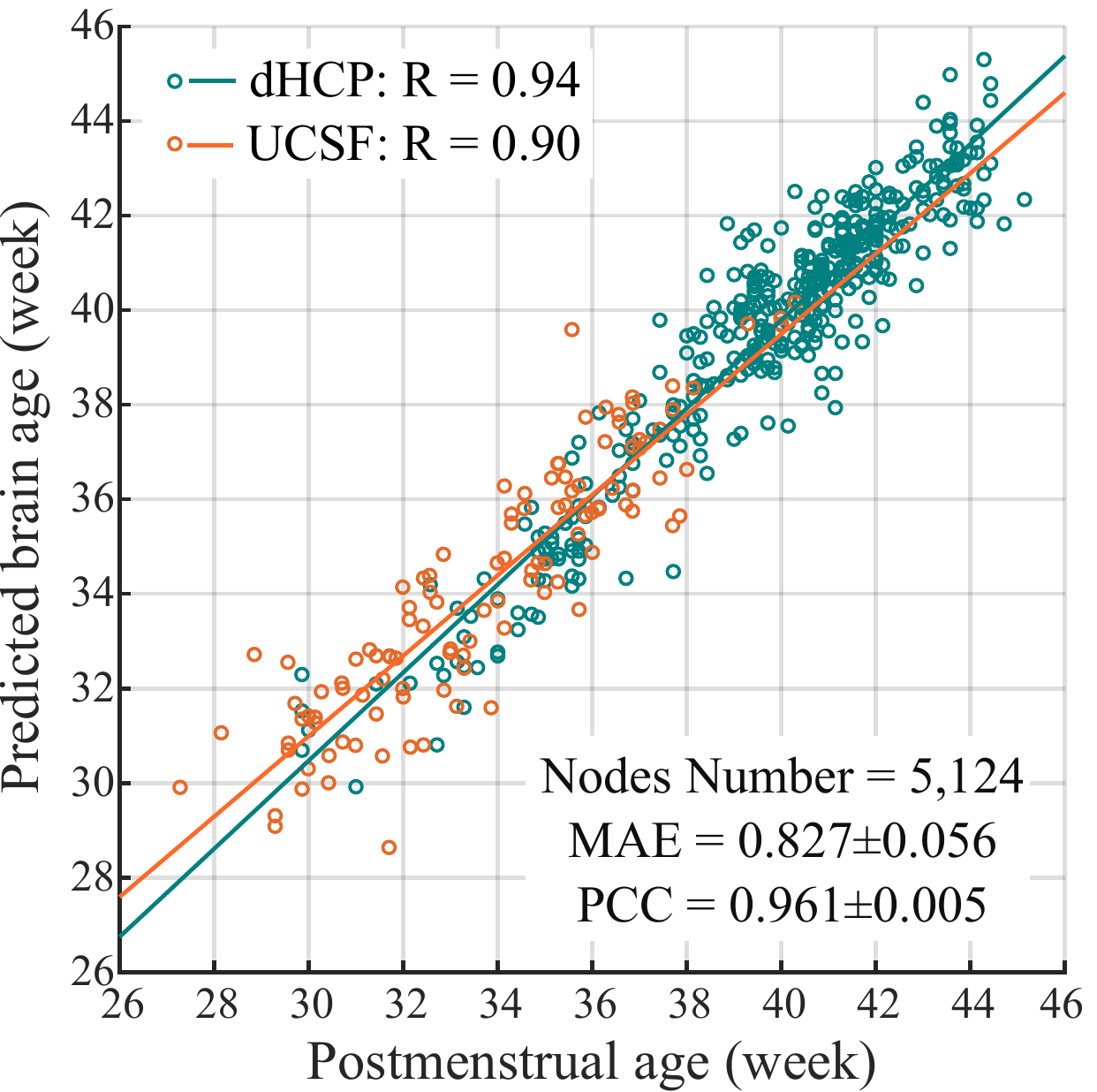}
	\caption{The scatter plot of the predicted brain ages and postmenstrual ages on the two cohorts with each input sparse graph consisting of 5,124 nodes. R: the correlation coefficient between the two axes for each cohort.}
	\label{fig6}
\end{figure}

Table \ref{table2} shows the performance of different models on the dataset, and several observations could be obtained: (1) Under identical hyperparameter, the SurfGNN model achieves optimal results when the input is sparse graphs with 5,124 nodes. Higher or lower resolution of input has worse MAE and PCC of results. Fig. \ref{fig6} presents the detailed prediction of SurfGNN with the sparse graphs input of 5,124 nodes, demonstrating a relatively consistent performance in two cohorts, albeit with slight deviation probably due to quantity imbalances. (2) SurfGNN consistently outperforms other algorithms across various sparse graph inputs without requiring the most parameters compared to its counterparts. Notably, when the node number is 5,124, our model exhibits the largest improvement in MAE compared to other models. (3) Higher input resolution and a larger number of nodes with cortical features do not always lead to better results. Although it can offer more detailed information, it may also bring something superfluous that troubles the models. Models of SphericalUNet, SiT as well as SurfGNN, have verified this. MoNet appears adept at distinguishing pertinent information from a larger node pool, but its overall prediction performance does not surpass that of our proposed model.

\begin{table}[!t]
	
	\caption{Performance comparison with different models on different resolution of the input sparse graphs.}
	\centering
	\renewcommand{\arraystretch}{1.1}
	\begin{threeparttable}
		\resizebox{\columnwidth}{!}{
			\begin{tabular}{cccccc}
				\toprule[1pt]
				\multicolumn{2}{c}{\multirow{2}{*}{Method}} & \multicolumn{4}{c}{\begin{tabular}[c]{@{}c@{}}Number of nodes of the input sparse graphs\end{tabular}} \\ \cline{3-6} 
				\multicolumn{2}{c}{}                        & 1,284                    & 5,124                   & 20,484                   & 81,924                   \\
				\midrule[1pt]
				\multirow{3}{*}{\begin{tabular}[c]{@{}c@{}}Spherical\\ UNet\end{tabular}}  & MAE        & 1.085$ \pm $0.059                         & 1.056$ \pm $0.125                         &1.036$ \pm $0.071                          & 1.058$ \pm $0.100                         \\
				& PCC        & 0.928$ \pm $0.011                         & 0.936$ \pm $0.019                        & 0.942$ \pm $0.004                         & 0.937$ \pm $0.013                         \\
				& Params (M) & 0.207                         & 0.897                        & 3.653                         & 6.802                         \\ \hline
				\multirow{3}{*}{MoNet}         & MAE        & 1.020$ \pm $0.042                         & 0.992$ \pm $0.024                        & 0.958$ \pm $0.025                         & 0.909$ \pm $0.015                         \\
				& PCC        & 0.944$ \pm $0.007                         & 0.946$ \pm $0.003                        & 0.951$ \pm $0.004                         & 0.955$ \pm $0.005                         \\
				& Params (M) & 0.230                         & 0.231                        & 0.239                         & 0.241                         \\ \hline
				\multirow{3}{*}{SiT}           & MAE        & 1.031$ \pm $0.054                         & 0.980$ \pm $0.032                        & 0.963$ \pm $0.023                         & 0.977$ \pm $0.005                         \\
				& PCC        & 0.944$ \pm $0.004                         & 0.948$ \pm $0.007                        & 0.950$ \pm $0.007                         & 0.950$ \pm $0.006                         \\
				& Params (M) & 7.172                         & 7.198                        & 7.260                         & 7.444                         \\ \hline
				\multirow{3}{*}{BrainGNN}      & MAE        & 1.047$ \pm $0.110                         & 1.026$ \pm $0.086                        & 0.964$ \pm $0.058                         & \multirow{3}{*}{\textbackslash}                         \\
				& PCC        & 0.941$ \pm $0.010                         & 0.944$ \pm $0.006                        & 0.945$ \pm $0.013                         &                          \\
				& Params (M) & 0.180                         & 0.241                        & 0.487                         &                          \\ \hline
				\multirow{3}{*}{SurfGNN}       & MAE        & 0.896$ \pm $0.075                         & \textbf{0.827$ \pm $0.056}                        & 0.835$ \pm $0.062                         & 0.887$ \pm $0.040                         \\
				& PCC        & 0.953$ \pm $0.010                         & \textbf{0.961$ \pm $0.005}                        & \textbf{0.961$ \pm $0.005}                         & 0.958$ \pm $0.005                         \\
				& Params (M) & 4.555                         & 4.568                        & 4.571                         & 4.609                         \\ \bottomrule[1pt]
			\end{tabular}
		}
		\begin{tablenotes}[para,flushleft]
			\item \textbackslash: results could not be obtained due to GPU memory limitations. \\
			\item M: millions
		\end{tablenotes}
	\end{threeparttable}
	\label{table2}
\end{table}

\subsection{Comparison on public datasets}
The achieved MAE of 0.827 in our model's results might not appear ideal when compared to similar studies\cite{b16,b18,b19} predicting brain age in neonates. This discrepancy could be attributed to a variety of factors, such as the use of varying datasets, distinct cortical surface reconstruction pipelines, the utilization of different cortical features for prediction, the selection of different hyperparameters in the models, etc. To validate the effectiveness and generality of our model, we apply it on a public dataset that is commonly used by several previous studies and compare its performance in state-of-the-art studies on PMA prediction. 

This dataset is derived exclusively from the dHCP project and includes data from 530 subjects from the dHCP dataset\cite{b38,b39,b40} . The extraction of the cortical surface and features is performed using the dHCP pipeline\cite{b35}. Four cortical features are utilized for prediction: sulcal depth, curvature, cortical thickness, and T1w/T2w myelination. Various studies have demonstrated the effectiveness of different models on this dataset. Notably, even when employing the same model, different studies have reported diverse results. In our study, we present findings from two published papers \cite{b18,b19} as well as results obtained from our own analysis. For a comprehensive evaluation, we compare our model's performance on this dataset against the top results reported in these papers, including those generated by our own analysis. The range of models compared includes geometric deep learning models such as SphericalUNet, MoNet, SiT, BrainGNN and MS-SiT.

Furthermore, we evaluate our model on a dataset of elderly individuals, comprising 1,000 subjects from the UK Biobank dataset (age range: 45.67-83.80 years, mean$ \pm $SD: 62.28$ \pm $6.84 years)\cite{b51}. We utilized CIVET\cite{b52}, an automated pipeline for structural human MRI, to extract cortical surfaces characterized by sulcal depth and cortical thickness. 

As depicted in Table \ref{table3}, SurfGNN demonstrates competitive performance on a public neonatal dataset, achieving a best MAE of 0.48. This result is close to, and in fact surpasses, the previously reported best MAE of 0.49. It highlights the generality and superior performance of SurfGNN compared to existing advanced models across various datasets. Additionally, our model employs significantly fewer parameters compared to optimal methods like SiT and MS-SiT. Moreover, as shown in Table \ref{table4}, SurfGNN achieves optimal prediction performance on a larger adult dataset, achieving a 4.289$\%$  improvement in MAE.

\begin{table}[]
	\caption{Performance comparison with different studies on the dHCP dataset}
	\centering
	\renewcommand{\arraystretch}{1.1}
	\begin{threeparttable}
		\resizebox{\columnwidth}{!}{
			\begin{tabular}{cccccc}
				\toprule[1pt]
				Method                                                   & MAE 1\tnote{1}    & MAE 2\tnote{2}    & MAE 3\tnote{3}    & Optimal MAE\tnote{4} & Params (M) \\ \hline
				\begin{tabular}[c]{@{}c@{}}SphericalUNet\end{tabular} & 0.57$ \pm $0.18 & 0.75$ \pm $0.18 & 0.69$ \pm $0.02 & 0.57$ \pm $0.18    & 2.64       \\
				MoNet                                                    & 0.57$ \pm $0.02 & 0.61$ \pm $0.04 & 0.68$ \pm $0.01 & 0.57$ \pm $0.02    & 0.31       \\
				SiT                                                      & 0.55$ \pm $0.04 & 0.54$ \pm $0.05 & 0.59$ \pm $0.01 & 0.54$ \pm $0.05    & 30.0       \\
				BrainGNN                                                   & \textbackslash        & \textbackslash & 0.76$ \pm $0.00        & 0.76$ \pm $0.00    & 0.30       \\
				MS-SiT                                                   & \textbackslash        & 0.49$ \pm $0.01 & \textbackslash        & 0.49$ \pm $0.01    & 22.0       \\ \hline
				SurfGNN                                                  & \textbackslash        & \textbackslash        & 0.48$ \pm $0.01 & \textbf {0.48$ \pm $0.01}    & 6.17       \\ \bottomrule[1pt]
			\end{tabular}
		}
		\begin{tablenotes}[para,flushleft]
			{\fontsize{8}{10}\selectfont \item \tnote{1}: performance of models reported in paper \cite{b18}.\\
			\item \tnote{2}: performance of models reported in paper \cite{b19}.\\
			\item \tnote{3}: performance of models achieved by us with the same data split. \\
			\item \tnote{4}: optimal performance of models observed in performance 1, 2 and 3.\\
			\item \textbackslash: results could not be obtained due to unreported in the papers or \\ the absence of code.\\
			\item M: millions}
		\end{tablenotes}
	\end{threeparttable}
	\label{table3}
\end{table}

\begin{table}[]
	\caption{Performance comparison on the UK Biobank dataset.}
	\centering
	\renewcommand{\arraystretch}{1.1}
	\begin{threeparttable}
		\resizebox{\columnwidth}{!}{
			\begin{tabular}{cccc}
				\toprule[1pt]
				Method        & MAE         & PCC           & Params(M) \\ \hline
				SphericalUNet & 4.014$ \pm $0.299 & 0.658$ \pm $0.060 & 2.179     \\
				MoNet         & 4.080$ \pm $0.199 & 0.636$ \pm $0.042 & 0.165     \\
				SiT           & 4.081$ \pm $0.390 & 0.639$ \pm $0.100 & 21.25     \\
				BrainGNN      & 3.915$ \pm $0.209 & 0.669$ \pm $0.064 & 0.154     \\ \hline
				SurfGNN       & \textbf{3.754$ \pm $0.106} & \textbf{0.699$ \pm $0.048} & 4.383     \\ \bottomrule[1pt]
			\end{tabular}
		}
		\begin{tablenotes}[para,flushleft]
			\item M: millions
		\end{tablenotes}
	\end{threeparttable}
	\label{table4}
\end{table}

\subsection{Ablation Study}
In this experiment, we conduct the ablation study to test the effectiveness of each module within the SurfGNN architecture using consistent hyperparameters and the sparse graph containing 5,124 nodes as input. Each component is tested as follows: (1) Multi-graph input. All three cortical features are integrated on a sparse graph for input. (2) Nodal graph convolutional layer. Replace it with the commonly used convolutional layer in Graph Attention Networks (GAT)\cite{b41}. (3) Score-weighted fusion mechanism. The score-weighted operation we designed (in \eqref{eq6}) is removed, and the other operations are preserved. (4) Other modules are tested by direct removal.

According to Table \ref{table5}, all modules of SurfGNN above contribute to the final result, with TSL, RSL and feature loss function emerging as the most effective components. This confirms the effectiveness of the graph analysis composed of TSL and RSL. Besides, the feature loss's advantage lies in imposing tighter constraints on parameter learning for each channel, effectively enhancing performance. Moreover, the specifically designed nodal graph convolutional layer appears more adept at characterizing cortical surface features compared to the GAT. The concept of enabling individual expression of cortical features within the model proves more effective than the strategy of collective feature extraction from amalgamated cortical features. Furthermore, our proposed fusion mechanism and score loss function, after rigorous testing, have been confirmed not to compromise the model's prediction performance.

\begin{table}[]
	\caption{Performance comparison on model without the corresponding module and the complete SurfGNN.}
	\centering
	\renewcommand{\arraystretch}{1.1}
	\resizebox{\columnwidth}{!}{
		\begin{tabular}{ccc}
			\toprule[1pt]
			Module                                                                                          & MAE                 & PCC                 \\ \hline
			Multi-graph input                                                                               & 0.956$ \pm $0.078          & 0.950$ \pm $0.009          \\
			Global graph convolutional layer                                                                & 0.986$ \pm $0.088          & 0.947$ \pm $0.008          \\
			Topology-preserved pooling layer                                                                & 0.981$ \pm $0.080          & 0.950$ \pm $0.008          \\
			Topology-sampling learning structure                                                            & 1.059$ \pm $0.075          & 0.938$ \pm $0.014          \\
			Nodal graph convolutional layer                                                                 & 1.024$ \pm $0.066          & 0.943$ \pm $0.009          \\
			Node-selective pooling layer                                                                    & 0.924$ \pm $0.041          & 0.951$ \pm $0.006          \\
			Region-specific learning structure                                                              & 1.005$ \pm $0.052          & 0.946$ \pm $0.010          \\
			Score-weighted fusion mechanism                                                                 & 0.875$ \pm $0.063          & 0.955$ \pm $0.008          \\
			Feature loss function                                                                           & 1.016$ \pm $0.077          & 0.943$ \pm $0.009         \\
			Score loss function                                                                             & 0.901$ \pm $0.033          & 0.955$ \pm $0.005         \\ \hline
			SurfGNN                                                                                         & \textbf{0.827$ \pm $0.056} & \textbf{0.961$ \pm $0.005} \\ \bottomrule[1pt]
		\end{tabular}
	}
	\label{table5}
\end{table}

\subsection{Number of TSL Structures}
The TSL structure has proven effective in our model, but the number of it within the model's framework is adaptable. The choice of resolution at which the surface topology begins with the RSL can influence the deeper graph analysis and the overall performance of the model.

\begin{figure}
	\centering
	\includegraphics[width=0.8\columnwidth]{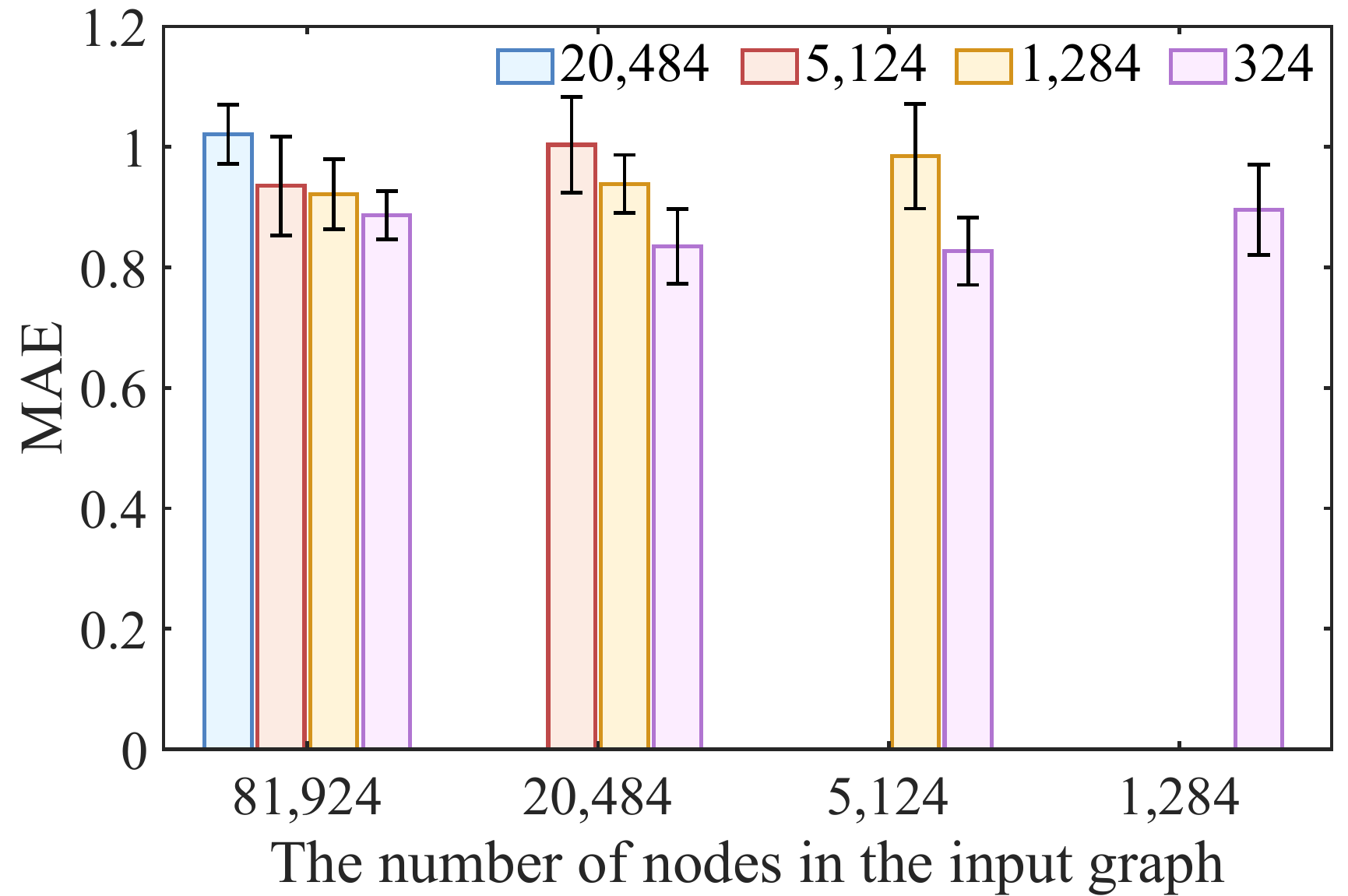}
	\caption{Comparison of prediction performance of SurfGNN containing different numbers of TSL structures, corresponding to distinct resolutions of output sparse graphs after all the TSL structures and also inputs of RSL structure.}
	\label{fig7}
\end{figure}

To access this, we conduct a comparative analysis to assess the prediction accuracy of input graphs subjected to varying degrees of topological condensation. Illustrated in Fig. \ref{fig7}, the horizontal axis indicates the resolution of the input sparse graph, and various colors represent the resolution of graphs where the RSL module begins. As expected, a clear pattern emerges: with more layers of TSL, the degree of topological condensation increases across inputs of differing resolutions, and the MAE decreases, signifying an improvement in prediction performance. This reinforces the crucial role of the TSL structure within the SurfGNN model. That is, for densely populated sparse inputs, a certain degree of down-sampling before in-depth graph learning significantly contributes to enhancing pertinent analysis.

\begin{figure*}[!t]
	\centering
	\includegraphics[width=0.8\textwidth]{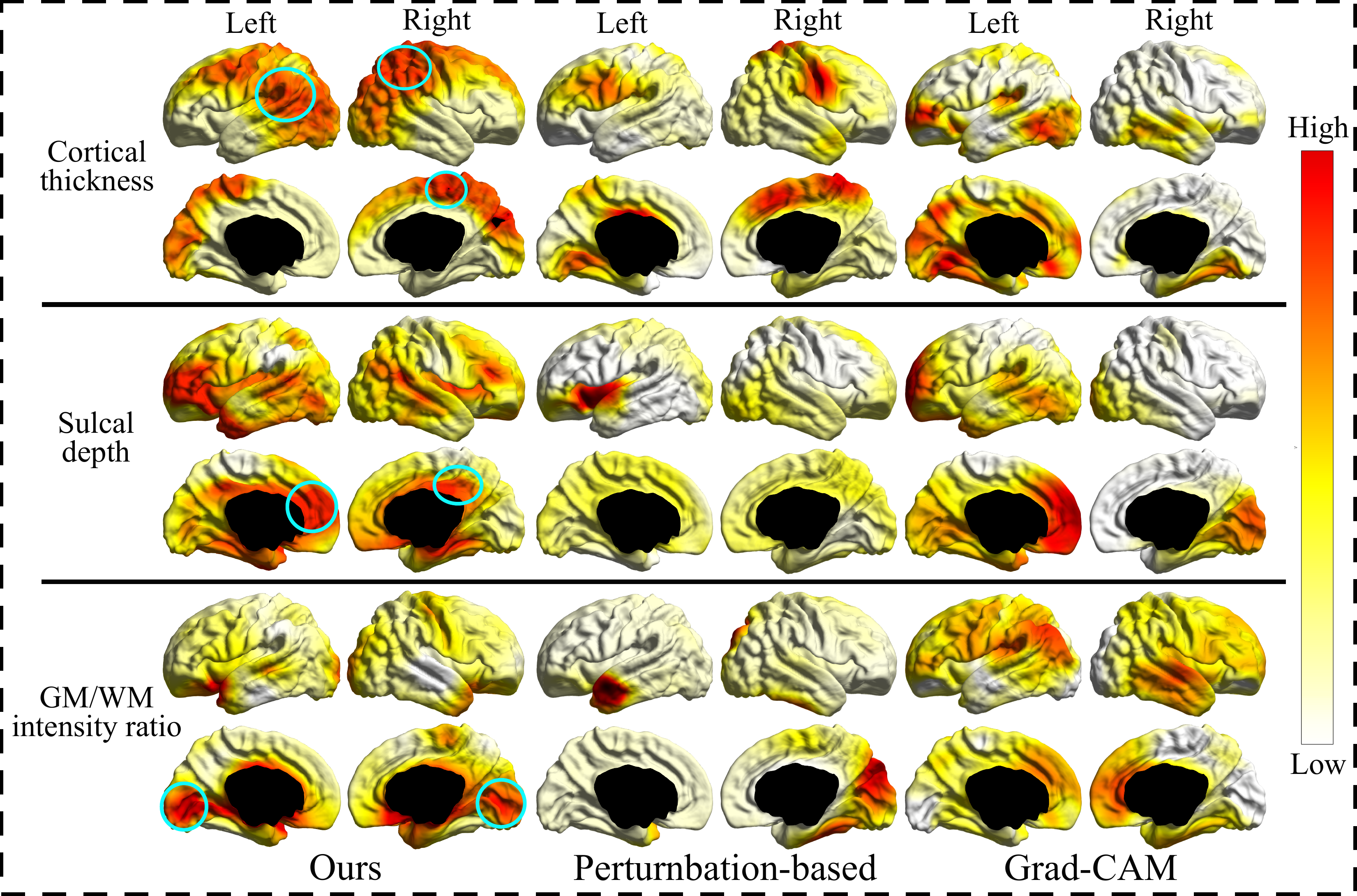}
	\caption{Comparison between our model and two post-hoc approaches on the spatial activation maps for the three cortical features. The circles indicate areas of higher response within each feature. The various colors for each surface denote differences in qualitative importance. Maps from different approaches or distinct cortical features of the same approach are not directly comparable in terms of values.}
	\label{fig8}
\end{figure*}

\begin{figure}
	\centering
	\includegraphics[width=0.7\columnwidth]{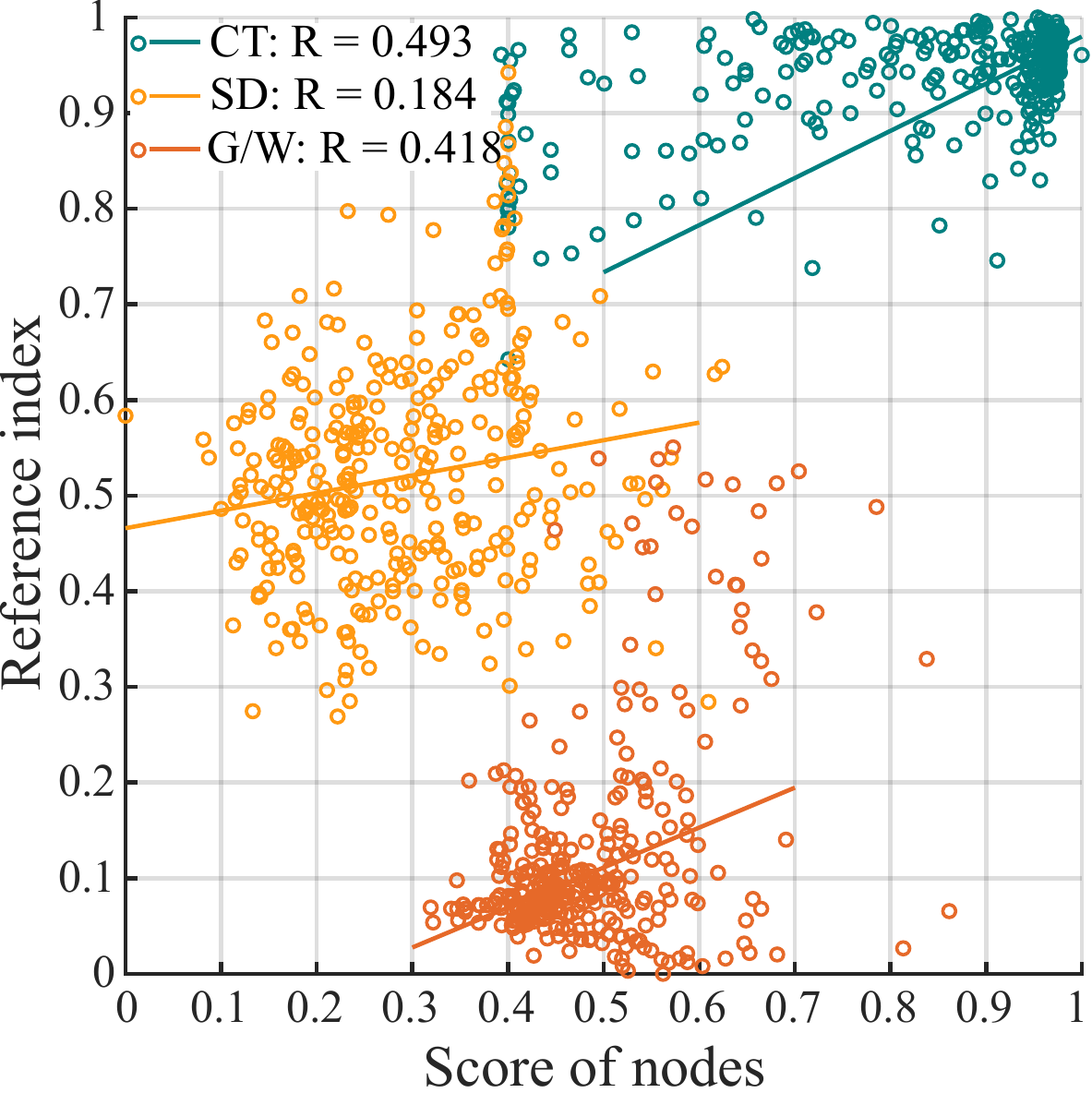}
	\caption{The scatter plots between the scores from model and the reference indexes on 324 nodes. CT: cortical thickness. SD: sulcal depth. G/W: GM/WM intensity ratio. R: the correlation coefficient between the two axes for each feature.}
	\label{fig9}
\end{figure}

\subsection{Interpretability}
SurfGNN has the capacity of independent interpretability for each cortical feature, via the score generated within the SWF structure. The node scores are mapped to the corresponding vertices in the brain surface, and after smoothing, a complete brain activation map is yielded, representing the spatial activation patterns for each cortical feature. Notably, in our model, after the TSL structures it consistently outputs a fixed down-sampled set of nodes, forming a sparse graph consisting of 324 nodes for each cortical feature. However, within the RSL structure, the decision of the node-selective pooling layers to drop nodes is dependent on learning and differs across subjects. Consequently, we obtain each activation map consisting of 324 nodes for each cortical feature (the score of nodes dropped in pooling layers is represented as 0). 

As Fig. \ref{fig8} shows, each cortical feature’s spatial activation map exhibits distinct regional heterogeneity, which varies across different features, indicating unique patterns of activation in response to prediction. In terms of physiological structure, the regions with higher activation for the feature of cortical thickness are the precentral gyrus and postcentral gyrus. For sulcal depth, it is the cingulate gyrus and cingulate sulcus, while another feature concentrates on the occipital lobe. Activation maps showcase the model’s capability in unveiling the influence of each node within each feature concerning the prediction results. In our task of predicting brain age, they illuminate the intricate relationship between different cortical features and neonatal brain development.

To evaluate the reliability of the spatial activation maps for each cortical feature, we introduce a reference index for each node score. This index is derived from the correlation coefficient between cortical feature measurements and ages for all nodes across the samples. The scatter plot in the Fig. \ref{fig9} illustrates node scores for each feature against their corresponding reference indexes. Across all nodes of the three cortical features, we observe a correlation of 0.580 between the node scores and reference indexes. Looking at individual feature, cortical thickness exhibits the highest correlation at 0.493, whereas sulcal depth shows the lowest at 0.184. Generally, there is a relatively consistent trend between node scores and reference indexes. Nodes with stronger age correlation tend to be more prominently represented in scores, as observed in the first feature. Conversely, Nodes with weaker correlation, as seen in the second and third features, exhibit lower scores.

Furthermore, the explanation performance of our SurfGNN model is evaluated in comparison to two representative feature-based explanation methods: the perturbation-based approach\cite{b53} and Grad-CAM\cite{b54}, both integrated with the MoNet model for post-hoc analysis. In the perturbation-based approach, if the input feature of key nodes in a well-trained model is replaced during the testing process, it could significantly impact the prediction results, which facilitates the analysis of each node’s contribution to the task. Grad-CAM operates by calculating the gradients in the last convolutional layer of the model, using these gradients as the activation response. For distinct activation maps of the three cortical features, we adapted the MoNet architecture to accommodate multi-graph input. Fig.\ref{fig8} illustrates the spatial activation maps produced by these two methods. To enhance readability, different colors are used to distinguish among regions. However, it is notable that the maps from different approaches or distinct cortical features of the same approach are not directly comparable in terms of values.

In addition to the correlation coefficient used as the reference index in Fig.\ref{fig9},  we introduce two additional metrics: sparsity and consistency. Sparsity measures the fraction of features deemed important by the explanation methods \cite{b55}. This metric is crucial, as effective explanations should be sparse, highlighting only the most significant input featrues while disregrading irrelevant ones \cite{b53}. It is calculated using the $ \ell_1 $ norm and $ \ell_2 $ norm for a overall evaluation. Furthermore, consistency assesses the stability of activation within the same age group. We posit that, for a given dataset, there should be a degree of similarity among subjects for explanations at comparable age stages. Specifically, the age range in the dataset, mainly spanning from 34 to 44 weeks, is divided into two-week intervals. We calculate the average standard deviation of activation values for all subjects within each group and use the average performance across all groups to evaluate consistency.  Table \ref{table6} presents the comparison of these three metrics among the interpretability methods. For the sparsity index, all three methods exhibit similar overall performance. However, in terms of relevance and consistency, our model significantly outperforms the other two methods. This indicates that our interpretable output is more aligned with the characteristic structure of the dataset and demonstrates a higher degree of stability within the groups.

\begin{table}[]
	\caption{Performance comparison with interpretability methods.}
	\centering
	\renewcommand{\arraystretch}{1.2}
	\resizebox{\columnwidth}{!}{
		\begin{tabular}{ccccc}
			\toprule[1pt]
			Method                            & Feature & Relevance      & Sparsity       & Consistency    \\ \hline
			\multirow{4}{*}{Pertubation-based} & CT      & 0.163          & 0.895          & 0.431          \\
			& SD      & 0.152          & \textbf{0.969} & 0.427          \\
			& G/W     & 0.263          & 0.859          & 0.346          \\
			& Total   & 0.087          & 0.860          & 0.401          \\ \hline
			\multirow{4}{*}{Grad-CAM}          & CT      & 0.002          & 0.885          & 0.298          \\
			& SD      & 0.068          & 0.920          & \textbf{0.270} \\
			& G/W     & 0.008          & 0.896          & 0.237          \\
			& Total   & 0.240          & 0.895          & 0.268          \\ \hline
			\multirow{4}{*}{Ours}              & CT      & \textbf{0.493} & \textbf{0.980} & \textbf{0.180} \\
			& SD      & \textbf{0.184} & 0.933          & 0.299          \\
			& G/W     & \textbf{0.418} & \textbf{0.985} & \textbf{0.188} \\
			& Total   & \textbf{0.580} & \textbf{0.898} & \textbf{0.223} \\ \bottomrule[1pt]
		\end{tabular}
	}
	\begin{tablenotes}[para,flushleft]
		\item CT: cortical thickness. SD: sulcal depth. 
		\item G/W: GM/WM intensity ratio.
	\end{tablenotes}
	\label{table6}
\end{table}

\section{Discussion}
\subsection{Model Composition}
In our work, we propose the SurfGNN, a specialized surface-based deep learning model tailored for phenotype prediction. We devise two learning modules, the TSL structure and the RSL structure, based on GNN for comprehensive surface analysis. Recent evidence suggests that the brain functions in a hierarchical manner that involves multiple scales\cite{b42,b43}. In this structure, information from lower, more localized levels is integrated and processed at higher, more global levels. The idea of our model aligns with this multi-scale framework. The TSL structure is primarily concerned with processing information at the lower scales, while the RSL structure deals with higher-scale organizational aspects.

The concept for the TSL structure originates from the spherical CNN\cite{b7, b11}, which transmits brain topology from densely connected, fine mesh at lower scales to a coarser mesh at higher scales. Despite its inherent weakness in detecting regional heterogeneities, it is highly effective in the initial learning and consolidation of complex graph structures. Conversely, the RSL structure effectively conducts in-depth graph analysis. Its convolution operation is tailored for specific vertices, offering an individualized approach for each vertex to assimilate higher-level information. The pooling operation in RSL is also region-specific, allowing it to select important vertices and exclude less significant ones from the prediction. The strength of the GNN-based information extraction structure lies in its ability to prioritize areas of the brain more closely associated with specific phenotypes, aligning with the natural workings of the brain, which is that the brain surface's response exhibits regional heterogeneity across various phenotypes. 

In short, the TSL module plays a crucial role in uniformly integrating local information from lower levels, ensuring that all information is unbiasedly transmitted to higher levels. Meanwhile, the RSL module is essential for region-wise feature manipulation and feature selection, emphasizing the most important information. In light of the brain's complex operations, a regional approach that favors a coarse-grained rather than a fine-grained analysis has proven to be more effective and practical in brain research\cite{b44,b45}. Our model expands on this concept by effectively incorporating local information, and refines the graph structure before proceeding with detailed graph analysis. The results in Table \ref{table5} underscore the indispensability of these two learning structures.

To further boost the predictive power of our model, we have innovated by allowing each cortical feature to develop its own distinct graph structure and a unique information extraction channel. Such a method is essential for comprehensively utilizing the key insights from all features that aid in prediction to improve the prediction accuracy, and it also plays a significant role in the feature-wise interpretations. Notably, the incorporation of independent channels in deep learning models usually results in a substantial increase in parameters and computational burden. However, for our model, this presents minimal challenges due to the limited number of cortical features typically utilized in brain surface analysis and the effectiveness of the two learning structures designed in SurfGNN. As indicated in Table \ref{table2} and Table \ref{table3}, it outperforms counterparts without necessitating the highest number of parameters, revealing its feasibility and effectiveness.

\subsection{Prediction Discrepancies in PMA Prediction from Different Datasets}
Table \ref{table2} and Table \ref{table3} present the outcomes of PMA prediction using identical models on distinct neonatal datasets, revealing a noticeable disparity between the two. Several factors contribute to this discrepancy. (1) Our dataset, although harmonized, amalgamates data from two cohorts, which produces dataset bias. (2) The UCSF cohort in our dataset contributes a higher proportion of preterm infant data than the public dataset. Typically, preterm infant subjects may exhibit health challenges, impacting overall prediction accuracy. (3) Our dataset encompasses a more diverse distribution of scan age (PMA: 27.28-45.14weeks, mean$\pm$std: 38.46$\pm$3.85 weeks) than the public dataset (PMA: 26.71-44.71 weeks, mean$\pm$std: 39.64$\pm$3.44 weeks). (4) The cortical feature extraction methods differ between the two datasets, with the NEOCIVET pipeline used for T1w MR images, and the dHCP pipeline relying on T2w MR images, respectively. Cortical features measured using different pipelines and modalities may yield variations in results, even for identical samples. (5) The selection of cortical features for prediction also influences results. The public dataset appears to have a more comprehensive selection, leading to more precise predictions.

\subsection{Input Resolutions and the Initial Resolution of RSL}
As indicated in Table \ref{table2}, our model consistently outperforms other algorithms across various sparse graph inputs, highlighting its robust adaptability. Nevertheless, inputs with varying resolutions also have a notable influence on prediction performance, a phenomenon observed across all models. Typically, the impact hinges on the learning capabilities of the models. When the input meshes consist of 1,284 nodes, all models exhibit poor prediction performance, likely due to the lack of crucial information in the lowest resolution graphs. For SurfGNN with 81,924 nodes input, a significant decrease in accuracy is observed. This may be attributed to an excessive number of graph convolutional layers during low-scale feature learning. SurfGNN utilizes a higher number of graph convolutional layers compared to other baseline models, increasing the risk of over-smoothing, which negatively impacts prediction performance. This issue is less prevalent in other baseline methods due to their simpler architectures with fewer graph convolution modules.

What’s more, as illustrated in Fig. \ref{fig7}, lower initial resolutions in RSL lead to better prediction performance. This observation can be attributed to several factors. Firstly, our model incorporates the TSL structure for initial learning and information  extraction from lower levels. Given the direct relevance of feature to our phenotype prediction task, having an adequate number of TSL structures for shallow graph analysis proves crucial. Additionally, within the RSL structure, employing concise input graphs alleviates the training burden on the model. Sparse graphs with fewer nodes facilitate the in-depth graph analysis by robustly capturing regions essential for prediction. Conversely, an abundance of nodes in the input sparse graph can lead to over-smoothing during learning, diminishing model performance.

\subsection{Reliability of Interpretability}
In section 5.5, we show the inherently interpretable results produced by the model, presenting spatial activation maps of prediction outcomes for different cortical features. In our task of predicting neonatal brain age, our objective for instance-level interpretability is to investigate the heterogeneous activities of cerebral cortex pertinent to neonatal brain development. Thus, metrics of relevance and consistency are crucial in the evaluation process. Our model demonstrates superior performance compared to the other two methods. In comparison with post-hoc approach, models with self-interpretability offer the advantage that effective activation responses during model training are typically more directly related to the output results, thereby influencing their accuracy. In our model, node scores are generated during the effective learning process from features to results, making them inherently aligned with the structural characteristics of the dataset. Furthermore, post-hoc analysis methods operate under the assumption that reliable explanations are contingent on accurate predictions. While these methods generally perform well with large-scale datasets, they are susceptible to being influenced by outliers in smaller datasets, making it challenging to achieve a consistent and effective interpretable output.

However, the interpretable result of our model has certain limitations, particularly in terms of relevance, as shown in Fig.\ref{fig9},  which can be attributed to several factors: (1) In our node-selective pooling layer, it's crucial to note that the dropped nodes are not necessarily irrelevant to the prediction. In our specific setup, a node is dropped if it is better represented by its neighboring nodes. Consequently, nodes relevant to predictions might be discarded due to their high similarity with information from surrounding nodes. They might contain relatively redundant information, rather than fundamentally unimportant information. Although this approach might result in greater prediction accuracy, it may be inconsistent with the interpretability. Relevant studies also have the similar conclusion that optimal prediction performance does not always align with optimal interpretable output\cite{b46}. (2) Node scores are computed though projected onto the same feature vector in SWF. Due to the limited feature length, it tends to yield a more parsimonious result, making it challenging to make precise comparisons. As Fig. \ref{fig9} shows, node importance for features of cortical thickness and GM/WM intensity ratio is concentrated in higher and lower areas, respectively, leading to a high correlation between the obtained node scores and reference indexes. Meanwhile, the importance distribution for the feature of sulcal depth is relatively scattered, resulting in a lower correlation. (3) For each cortical feature, our initial input consists of 5,124 nodes. In other words, each activation map of 324 nodes we obtain integrates the features of adjacent nodes at a higher level of resolution, rather than solely representing each node’s features. In the visualization of activation maps, we have addressed this issue with smoothing. However, our reference index only calculates the features of a single node, causing a slight misalignment in meaning between the reference index and the node score.
\subsection{Limitations}
Limitations of our work and future directions: (1) Our SurfGNN architecture, particularly the TSL structure, emphasizes local direction-invariant features for efficient feature extraction and topological fusion. However, it may inadvertently overlook local shape features, leading to insufficient low-scale feature extraction when high-resolution inputs are utilized. This limitation potentially reduces the overall performance of the model. Future work will focus on developing a more robust and comprehension information extraction module to address the shortcomings and enhance efficacy. (2) Whereas our model has demonstrated promising performance on three different datasets,  larger and various datasets covering a full age spectrum should be utilized in the future. (3) Extending beyond the brain age prediction, the next step involves applying our model, built on a healthy cohort to cohorts with neurological disorders. This will allow us to investigate phenotypic distinctions and spatial differences in cortical features for individuals with abnormalities. (4) The precision of the interpretable output is subject to limitations. Investigating interpretability that is more precise and well balanced with prediction accuracy could be the direction for improvement.

\section{Conclusion}
In conclusion, we have presented a novel graph neural network for phenotype prediction utilizing cortical surfaces. Specifically, our approach incorporates TSL and RSL structures tailored for graph analysis across different resolutions of surface mesh. Additionally, the design of multi-graph input and the score-weighted fusion mechanism provide performance optimization and effective interpretability for prediction tasks. In neonatal brain age prediction, the model has demonstrated superior performance alongside competitive interpretable outputs. Future works encompass validation on larger datasets, extension to diseased cohorts, exploring models that strike a balance between prediction and interpretability, and enhancing the precision of interpretable results.

% Uncomment and use as the case may be
%\begin{theorem} 
%\end{theorem}

% Uncomment and use as the case may be
%\begin{lemma} 
%\end{lemma}


% \bibliographystyle{cas-model2-names}




\end{document}